\documentclass[onecolumn,fleqn,10pt]{wlscirep}
\usepackage[utf8]{inputenc}
\usepackage[T1]{fontenc}
\usepackage{bm}
\usepackage{amsmath}
\usepackage{amsthm}
\usepackage{amssymb}
\usepackage{ulem}
\usepackage{mathptmx}
\usepackage{mathrsfs}
\usepackage{gensymb}
\usepackage{physics}
\usepackage{scicite}
\usepackage{times}
\usepackage{lineno}
\usepackage{soul}
\usepackage{comment}
\usepackage{rotating} 
\usepackage{adjustbox}
\usepackage{pdflscape}  
\usepackage[mathscr]{eucal}
\usepackage{eucal}

\title{Manipulating ferroelectric topological polar structures with twisted light}

\author[1]{Nimish P. Nazirkar}
\author[2,\dag]{Viet Tran}
\author[2,\dag]{Pascal Bass\`ene}
\author[1]{Atoumane Ndiaye }
\author[2]{Julie Barringer}
\author[1]{Jie Jiang}
\author[3]{Wonsuk Cha}
\author[3]{Ross Harder}
\author[1]{Jian Shi}
\author[2,\dag,*]{Moussa N'Gom}
\author[1,2,*]{Edwin Fohtung}
\affil[1]{Department of Materials Science and Engineering, Rensselaer Polytechnic Institute (RPI), Troy, NY~12180, USA}
\affil[2]{Department of Physics, Applied Physics, and Astronomy, RPI, Troy, NY~12180, USA.}
\affil[3]{Advanced Photon Source, Argonne National Laboratory, Lemont, Illinois 60439, USA}
\affil[$\dag$]{Shirley Ann Jackson, Ph.D. Center for Biotechnology and Interdisciplinary Studies, Rensselaer Polytechnic Institute, Troy, New York 12180, USA.}
\affil[*]{e-mail:ngomm@rpi.edu and fohtue@rpi.edu}

\begin{abstract}

The dynamic control of novel ordered states of matter, particularly those unattainable in thermodynamic equilibrium, remains a cornerstone of condensed matter physics. Notably, intense terahertz fields have been used to induce metal-insulator transitions, superconductivity, ferroelectricity in quantum paraelectrics, and room-temperature magnetization through circularly polarized terahertz electric fields\cite{von2018probing, liu2012terahertz, fausti2011light, li2019terahertz, basini2022terahertz}. Central to these phenomena is the excitation of infrared-active soft phonon modes by terahertz electric fields. Extending this paradigm, recent theoretical work suggests that ferroelectric polarization can be manipulated using terahertz twisted light, which transfers orbital angular momentum to induce ferroelectric skyrmions\cite{gao2024dynamical,gao2024effective}.

In this study, we present experimental evidence demonstrating that such control is achievable in the quasi-2D ferroelectric CsBiNb$_2$O$_7$ using twisted ultraviolet (UV) light carrying orbital angular momentum (OAM). By resonantly exciting both the zone-center ferroelectric mode and the zone-boundary octahedral tilting mode, we utilize twisted UV light to dynamically modulate the ferroelectric polarization. We develop and employ in-situ X-ray Bragg coherent diffractive imaging, twisted optical Raman spectroscopy, and density functional theory calculations to detect and three-dimensionally resolve the resulting ionic displacement fields and changes in ferroelectric polarization textures.

Our observations reveal deterministic and reversible twisted light-induced strain and ionic displacements within the unit cell, leading to significant microscopic changes in ferroelectric polarization. The interplay between twisted optical photons, phonon modes, and induced ionic displacements reduces the material’s symmetry, thereby stabilizing a non-equilibrium ferroelectric phase that harbors topological solitons far from thermodynamic equilibrium. These findings pave a new path for controlling ferroelectricity—and potentially magnetism. This research opens avenues for developing novel optoelectronic devices, such as ultrafast non-volatile memory switches, by harnessing light to coherently control ferroic states.

\end{abstract}

\begin{document}

\flushbottom
\maketitle

\noindent \textbf{Key points:}: Twisted light to control polarizations in quasi-two-dimensional FE materials,Bragg coherent X-ray diffractive imaging, data storage, information processing, and energy harvesting

\section*{MAIN} 


In recent years, there has been a surge in condensed matter research towards nontrivial topologically-protected dipolar textures. These textures have predominantly been found in bulk materials, nanocrystals, and epitaxial heterostructures, where precise electrostatic, magnetostatic, and elastic boundary conditions are enforced on crystalline ferroelectric (FE) and magnetic materials. Since the prediction of flux-closure domain structures in magnetic films in the 1940s~\cite{kittel1949physical,kittel1946theory}, parallels have been drawn in predicting and observing real-space topologically-protected dipolar textures such as flux-closures~\cite{tang2015observation}, center domains~\cite{ma2018controllable}, vortices~\cite{karpov2017three,yadav2016observation}, skyrmions~\cite{das2019observation}, merons~\cite{wang2020polar},  bubbles~\cite{zhang2017nanoscale}, and hopfions~\cite{luk2020hopfions} in FEs.

The typically smaller sizes of these FE domains, compared to their magnetic counterparts, offer promising applications in high-density information storage, thin-film capacitors, actuators, and other electronic devices~\cite{kittel1946theory,kittel1949physical,wang2020polar}. Recent theoretical and experimental studies have indicated the presence of Bloch points (BPs) in several polar systems. For instance, Salje and Scott found Bloch lines and BPs within polar domain walls in SrTiO$_3$~\cite{salje2014FE}, while Morozovska et al.~\cite{morozovska2020electric} observed BPs in BaTiO$_3$ nanoparticles with SrTiO$_3$ shells under certain conditions. Gao et al.~\cite{gao2024dynamical} demonstrated that an optical vortex beam could generate skyrmion states in ultrathin Pb(Zr$_{0.4}$Ti$_{0.6}$)O$_3$ films, leading to the formation of BPs. Despite these advancements, experimentally observing and manipulating topological textures such as vortices, merons and BPs in FE systems remain challenging as they all require strict symmetry constraints and control of the vectorial nature of the applied electric field.

In our earlier work, we demonstrated the volumetric observation of ferroelectric (FE) domains and polar vortices in isolated BaTiO$_3$ nanoparticles~\cite{karpov2017three,karpov2019nanoscale,shi2023enhanced}. Additionally, other studies have experimentally observed phenomena such as negative capacitance and the emergence of chirality in polar vortices and skyrmions, originating from sequences of achiral materials, using resonant soft X-ray diffraction-based circular dichroism and optical measurements~\cite{behera2022electric, shafer2018emergent, lovesey2018resonant, chauleau2020electric}. Recent advances, including phase field simulations~\cite{wang2024polar} and aberration-corrected scanning transmission electron microscopy, have further shown that polar BPs can be stabilized in substrate-induced tensile-strained ultrathin FE PbTiO$_3$ films.

Despite these significant advancements, the experimental observation and manipulation of polar textures, particularly in a nondestructive manner and in three dimensions, remain challenging endeavors~\cite{Jeong2024,karpov2017three}. Furthermore, the ability to apply inhomogeneous electric fields without the use of electrodes, while simultaneously imparting strain inhomogeneities (e.g., tensile, compressive, twist), has proven elusive. Traditionally, the control, stabilization, and manipulation of such FE topological structures have been achieved through substrate-induced strain in ferroelectric/dielectric superlattices~\cite{yadav2016observation,das2019observation,behera2022electric} or epitaxial thin films~\cite{wang2024polar,wang2020polar} under external electric fields. However, these methods face significant challenges, as they are constrained by mechanical boundary conditions rigidly defined by the atomic structure of the substrate or proximity layer. Moreover, manipulating these topological structures often requires the use of AFM contact tips as gates or the design and fabrication of electrodes with high precision to apply nonlinear electric fields at nanoscale resolutions, which is technically demanding.

In this work, we address these challenges by experimentally demonstrating the stabilization of polar BPs due to TL-induced tensile strain in 2D FE flakes using coherent diffractive imaging. We observed the internal 3D structures of these BPs and their reversible transition from BPs to merons and back to BPs. Moreover, we revealed the phenomenon of inducing dynamical steady-state strain inhomogeneities. Our findings not only advance the fundamental understanding of topological textures in FEs but also open new pathways for their application in next-generation ultrafast technologies.

\begin{figure}[!t]
\centering
\includegraphics[width=0.95\textwidth]{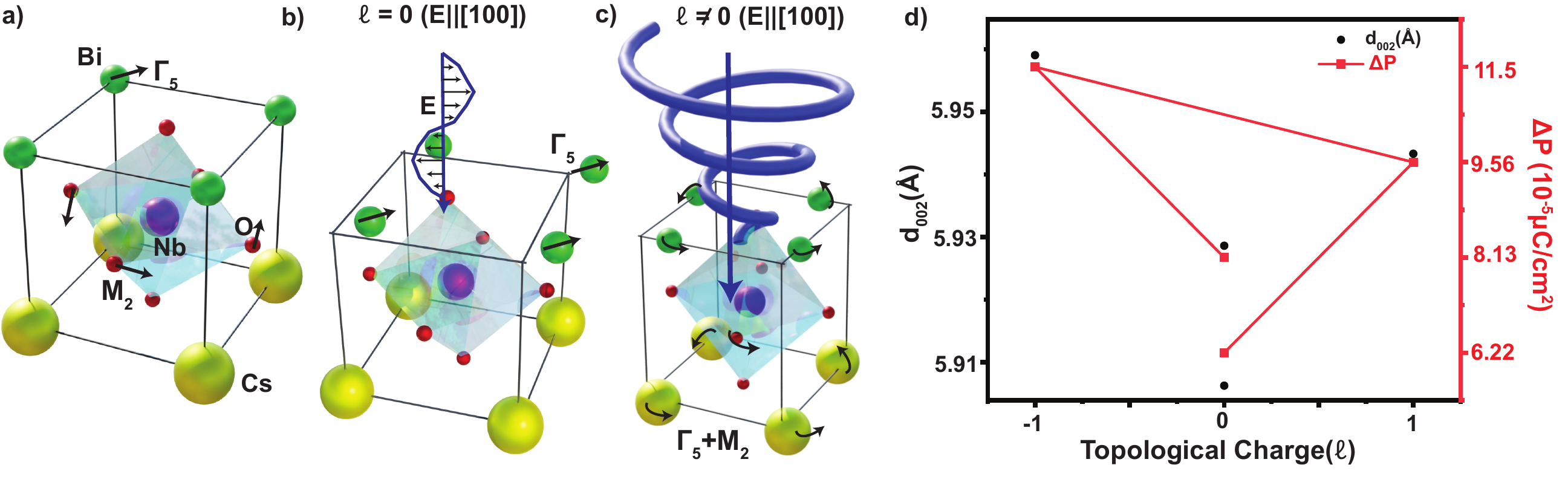}
\caption{\textbf{Schematic of the manipulation of FE using vortex field from twisted light and its experimental detection.}\\ \textbf{(a)} The CsBiNb$_2$O$_7$ crystal unit responsible for FE, shown in the absence of twisted light. When driven by a plane-polarized 375 nm UV field in a Laguerre-Gaussian (LG$_\ell$) mode carrying orbital angular momentum (OAM) quantized at $\ell\hslash$ per photon, it induces orbital atomic motions around the beam's propagation axis. This motion creates a net dipole moment per unit cell and lowers symmetry through resonant excitation of both the zone-center and the zone-boundary octahedral tilting, FE modes. \textbf{(b, c)} Twisted coherent UV beam topological charges \(\ell = 0\) and \(\ell \neq 0\) drive the zone-center \textbf{(b)} and the zone-boundary octahedral tilting \textbf{(c)} modes, respectively. \textbf{(d)} Experimentally observed changes in polarization and atomic displacements along the [002] direction in CBNO arising due to the phonon modes depicted in \textbf{(b~\&~c)}. }
\label{fig:1}
\end{figure}

It has been established theoretically~\cite{gao2024dynamical,gao2024effective} that TL in the THz regime can excite and drive dynamical polar skyrmions and transfer its nonzero winding number to FE ultrathin Pb(Zr\(_x\)Ti\(_{1-x}\))O\(_3\) films. This interplay between the orbital angular momentum (OAM)~\cite{allen1992orbital} of TL and electric dipoles in FE material leads to a distinctive alignment within the polarization fields. While the TL electric field \(\bm{E}_{\text{OAM}}\) exhibits uniformity across layers, the \(z\)-component of the polarization exhibits layer-dependent variations, contrasting with the consistent in-plane components. The crux of this interaction is the electrostatic coupling \(-\bm{p} \cdot \bm{E}_{\text{OAM}}\), which governs the energetically favorable orientation of the dipoles leading to changes in the out-of-plane polarization \(P_z\). This subtle alignment, in turn, drives the redistribution of bound charges, tailoring the \(z\)-component of the electric dipoles and hence, the evolution of polar skyrmions. Theoretical predictions suggest this coupling can induce polarization modulations ranging from 0.2 to -1.0 C/m\(^2\) in response to targeted ionic displacements within the PZT films. Inducing and detecting such ionic motion in an isolated free-standing quasi-two-dimensional FE material is the main idea of the experiment presented here and illustrated in Fig.\ref{fig:1}.

 Layered perovskite such as the Dion-Jacobson (DJ) phase CsBiNb$_2$O$_7$ (CBNO) are not only a significant class of FEs but also provide an ideal system for exploring the mechanisms that induce polar structures, especially under OAM illumination. These mechanisms include second-order Jahn-Teller-driven cation displacements, cation ordering patterns, and combinations of nonpolar octahedral rotations. CBNO is an improper FE with \textit{P2$_1$am} symmetry at room temperature~\cite{fang2022hybrid}. It is particularly suited for probing electric-field-driven dynamics due to its several zone-center phonon modes within the frequency range accessible to modern terahertz sources. As illustrated in Fig.~\ref{fig:1}(a), theory suggests that the interplay between $\Gamma$-point zone-center phonons and M-point zone-boundary phonons in CBNO drives its critical phase transitions~\cite{fennie2006FE}. Specifically, the transition from \textit{P4/mmm} to \textit{Pmam} is associated with the stabilization of an M$_5$ zone-boundary phonon, while the transition to P2$_1$am symmetry requires the cooperative action of a $\Gamma_5$ zone-center FE mode and an M-point zone-boundary mode, accounting for the observed symmetry change. Surface-sensitive experiments~\cite{fang2022hybrid} and density functional theory (DFT)  predict significant FE behavior in the DJ phase of CBNO, with a polarization \( \bm{P} \approx 30-40 \, \mu\text{C cm}^{-2} \), much higher than that in the $n=2$ Aurivillius phase SrBi$_2$Ta$_2$O$_9$~\cite{fennie2006FE}. Structural studies also confirm that the adopted room temperature polar, orthorhombic structure (\textit{P$2_1am$} symmetry, $\sqrt{2a_t} \times \sqrt{2a_t} \times c_t$) is stable up to at least 1000 $^\circ$C~\cite{chen2015FE}. This polar structure originates from a tetragonal structure through the rotation of NbO$_6$ octahedra around both an in-plane axis ([110]$_t$, $a^-a^-c^0$ in Glazer notation, corresponding to the (M$_{5^-}$) rotational mode) or the tetragonal axis [001]$_t$ ($a^0a^0c^+$ , corresponding to the (M$_{2^+}$) rotational mode), along with the FE displacement of Bi$^{3+}$ cations relative to its neighboring O$^{2-}$ - ions along the polar $a-$axis~\cite{goff2009leakage, snedden2003structural} as depicted in Fig.~\ref{fig:1}(a). This structure is analogous to FE SrBi$_2$Ta$_2$O$_9$ and related off-stoichiometric phases~\cite{hervoches2001two}.

Our recent experimental investigations into pyroelectric measurements using both HAADF-STEM and XRD have demonstrated the presence of giant pyroelectricity\cite{jiang2022giant} and room-temperature FE with in-plane domain in free-standing quasi-two-dimensional CBNO flakes\cite{guo2021unit}. However, experimental operando demonstration of its three-dimensional displacement fields and domain structure due to FE behavior has been elusive until now~\cite{goff2009leakage, snedden2003structural}.

Moreover, distinguishing between types and volumetric probing of topological FE polar structures and solitons and 3D induced strain under OAM illumination (Fig.~\ref{fig:1}(b-c)) necessitates the development of non-destructive, high spatial resolution probes, namely Bragg coherent diffractive imaging (BCDI)~\cite{karpov2019nanoscale, robinson2009coherent} capable of tracking changes in inter-atomic spacing and microscopically induced polarization (Fig.~\ref{fig:1}(d)) in a quazi-2D free-standing FE crystals. These changes in atomic spacings are in line with a previously reported study on temperature-dependent analysis in CBNO~\cite{Goff2009}.

In this article, we present experimental evidence demonstrating the manipulation of FE and tracking  d$_{002}$ spacing by combining \textit{in-situ} 375~nm continuum twisted UV light, generated with a tabletop source, with X-ray BCDI~\cite{karpov2017three,robinson2009coherent} (see Fig.~\ref{fig:2}(a) and in \textbf{Materials and Methods}) and optical Raman spectroscopy. We use BCDI to reconstruct 3D atomic displacements and strain maps of CBNO nanoflakes and gain an understanding of the unusual TL control of FE polar structures. Theoretical computations are consistent with these observations and provide additional insights into the origin of improper FE in 2D Dion–Jacobson systems. Our observations show FE domains and $\mathrm{Z}_{2} \times \mathrm{Z}_{4}$ polar vortex loops spanning the volume of CBNO nanocrystals, suggesting the use of CBNO as an alternative to rare-earth manganites for studies of the Kibble-Zurek mechanism in condensed matter. By varying the topological charge (see Fig.~\ref{fig:2}(c)) of TL, we can manipulate FE and study the dynamics of FE domains, vortex loops, and the vorticity (winding number) of FE polarization near the loops. These loops appear to be stable and locally pinned due to the topologically protected FE vortices. We identify four antiphase FE domains with left/right in-plane polarization configurations that form a vortex or an antivortex. Our results suggest that TL can be used to manipulate FE and create stable, non-trivial FE topological polarization textures associated with the FE displacement field.

\section*{RESULTS}

To conduct the \textit{in-situ} TL BCDI experiment, we first synthesized free-standing CBNO 2D nanoflakes using a polymer precursor method~\cite{guo2021unit}. These nanoflakes were then transferred onto the sample holder (see \textbf{Materials and Methods} and {\bf Supplemental Data}, Figs. S1 and S2).

TL, also known as optical vortices or light carrying OAM, consists of electromagnetic fields~\cite{allen1992orbital,forbes2017controlling} with one or more singularities where the phase~\cite{allen1992orbital} is undefined, causing the intensity to vanish. The discovery of TL beams in visible light produced in free-space~\cite{allen1992orbital,wang2023free} has generated significant interest across various scientific disciplines, including quantum information processing~\cite{erhard2018twisted}, optical tweezers, super-resolution microscopy~\cite{shen2019optical}, and telecommunications~\cite{tran2023exploration}.

\begin{figure}[!th]
\centering
\includegraphics[width=.95\textwidth]{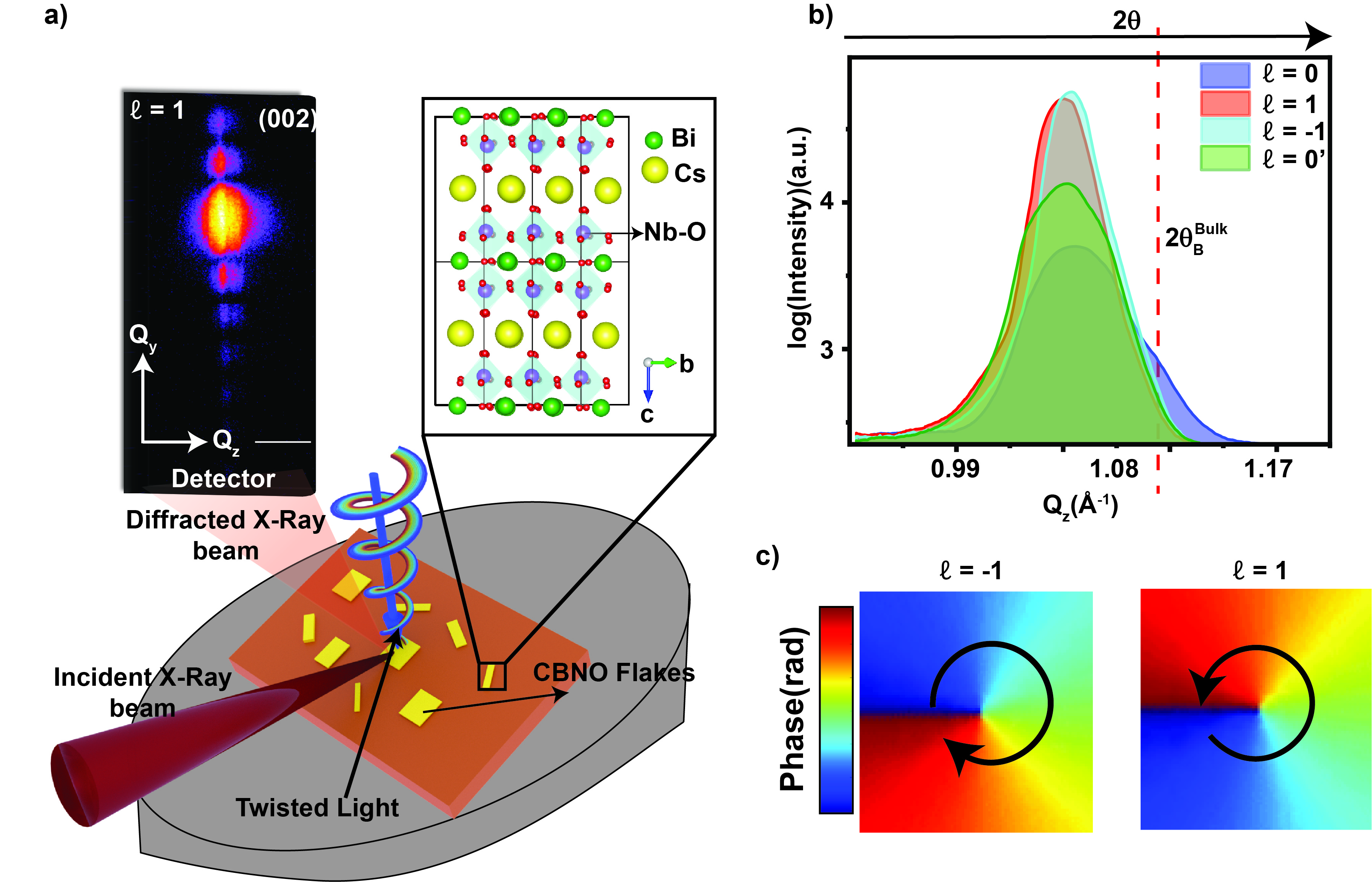}
\caption{\textbf{In-situ Bragg coherent diffractive imaging  experiment}. 
\textbf{(a)} Coherent x-rays (red) are incident on a nanoflake (yellow) containing polarization vortices and FE domains subjected to twisted UV light (Red-yellow-blue) illumination. 
A schematic of a Dion-Jacobson layered oxide CBNO crystal which has been shown to harbor FE with high Curie temperature and large in-plane polarization is used as a model system to study twisted light control of atomic displacements, microscopic FE polarization and domain structure with near-atomic resolution observed as 
\textbf{(b)} shifts and broadening of the recorded Bragg peak intensity.  
Simulated phase of twisted UV lights with topological charges ($\ell$) of \(1\) and \(-1\), carrying OAM quantized in units of \(\ell\hslash\) per photon 
\textbf{(c)}. In our experiments, the diffracted X-rays collected at the detector plane carry information about the 3D electron density and atomic displacement fields within the nanoflake that allows us to monitor recorded X-ray intensity variations near the Bragg spot during a cyclical application of light topological charge from \(\ell = 0\), \(1\), \(-1\), and back to \(0\).}
\label{fig:2}
\end{figure}

Figure~\ref{fig:2}(c) illustrates exemplary phase and 2D intensity profiles (see {\bf Supplemental Data}, Figs.~S4 and S29) in planes perpendicular to the propagation direction ($z$-direction) for the Laguerre-Gaussian (LG$_\ell$) mode of TL used in this experiment. LG beams are solutions to the paraxial wave equation~\cite{allen1992orbital}, and in the lowest order of the paraxial approximation, the electric and magnetic fields, as well as the vector potential, are purely transverse. Using the Lorenz gauge, the longitudinal component of the electric field (see {\bf Supplemental Data}, S7 and Figs. S4, S5, S6, and S7) can be expressed in terms of a scalar potential as:
\begin{equation}
E_z(\mathbf{r}, t) = -\partial_t \Phi(\mathbf{r}, t) = -c e^{i(kz - \omega t)} \mathbf{A}_0 \cdot \nabla_\perp \Pi(\mathbf{r}),
\end{equation}
where $E_z$ is the longitudinal component of the electric field, $\Phi$ is the scalar potential, $c$ is the speed of light, $k$ is the wave number, $\omega$ is the angular frequency, $\mathbf{A}_0$ is the vector potential amplitude, $\nabla_\perp$ is the transverse gradient, and $\Pi(\mathbf{r})$ is the complex amplitude of the LG beam, expressed as $\Pi(\mathbf{r}) = \Pi_{0}(r,z)\exp(i\ell\phi)$.

In our \textit{in-situ} BCDI setup, we utilized a spatial light modulator (SLM) (more details in \textbf{Materials and Methods}) to focus a 375 nm continuum ( with constant beam power over time) of LG beams with varied topological charge $\ell$ normal to the in-plane direction ([100]-direction) of our CBNO sample (see Fig. \ref{fig:1}). We generated a spatially inhomogeneous TL electric field $\textbf{E}_{\text{OAM}}$ (\ref{EOAM}) and measured its intensity profile. For full derivation, see {\bf Supplementary Data}, Figs. S4, S5, and S6 and Note S7.
\begin{equation} \label{EOAM} 
\textbf{E}_{\text{OAM}}(\mathbf{\rho}, t) = 
\begin{cases} 
E \left( \frac{\sqrt{2}\rho}{w} \right)^{|\ell|} e^{-\frac{\rho^2}{w^2}} \left( \cos(\ell\phi - \omega t) \mathbf{e}_x - \sigma \sin(\ell\phi - \omega t) \mathbf{e}_y \right), & \text{for } 0 \leq t \leq T, \\
0, & \text{for } t > T.
\end{cases}
\end{equation}

\begin{figure}[!t] 
\centering
\includegraphics[width=0.95\textwidth]{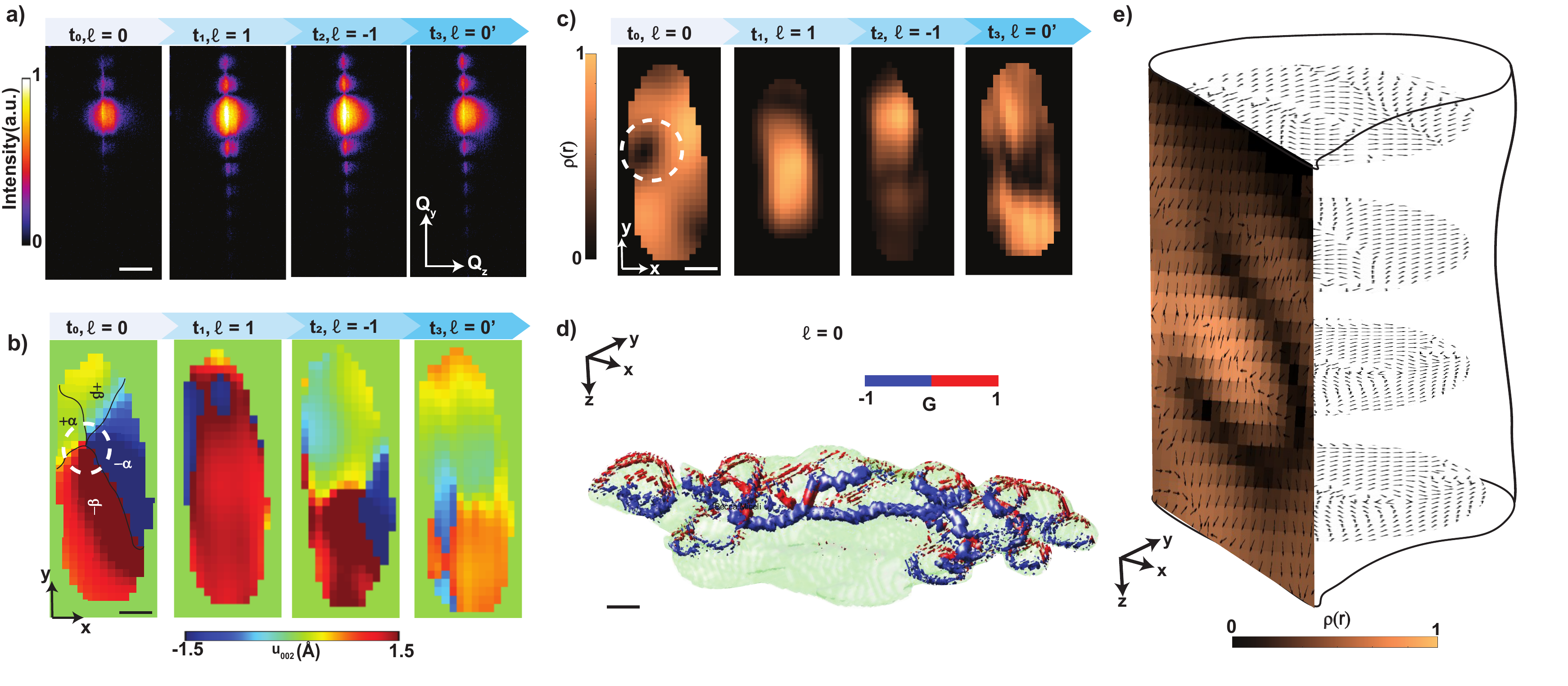}
\caption{\textbf{Real-space topologically-protected polar structures under a vortex-light field.} 
\textbf{(a)} Two-dimensional slices from the recorded three-dimensional coherent diffraction patterns for the $(002)$ Bragg peak, used to measure the time evolution of the \textbf{(b)} FE displacement field during twisted light-induced structural changes. The polarization vortex core is identified at time $t_{0}$ under LG$_\ell$ beam illumination of $\ell = 0$ (zero OAM). As time progresses to $t_{3}$, changes in the displacement field are observed due to the nucleation of a vortex-antivortex pair. \textbf{(c)} A two-dimensional map showing the evolution of the central cut of a Bloch point in the freestanding CBNO. \textbf{(d)} The magnitude of the toroidal moment density under LG$_\ell$ beam of $\ell = 0$. \textbf{(e)} A rendering at time $t_{0}$ of CBNO under a TL electric field of $\ell = 0$, illustrating a vortex spanning the volume of the crystal along the $z-$direction. Each cross-section perpendicular to the $ z-$ direction contains a vortex texture, and the cut through the $y-$ direction shows a bloch-like behavior around the vortex core. The scale bar in \textbf{(a)} represents 0.01 \(\text{\AA}^{-1}\) while that in {\bf (b), (c),} and {\bf (d)} represents ~60 nm.}
\label{fig:3}
\end{figure}

In equation (\ref{EOAM}), $T$ is the time during which the CBNO nanoflake is under LG$_\ell$ beam illumination. 
At the time $t = 0$, we switched on the TL UV beam and, with the aid of the SLM and the neutral density filter, selected TL electric field polarization($\sigma = 0$) components $\mathbf{e}_x$ and $\mathbf{e}_y$ and magnitude such that {\bf E}$_\parallel [100]$ as shown in Fig. \ref{fig:1}. We aligned the X-ray detector and CBNO crystal to satisfy the Bragg condition for the $d_{002}$ inter-atomic spacing of the (002) Bragg spot, as shown in Fig. \ref{fig:3}. For a duration $T$, we measured 3D diffraction near the (002) peak for beam topological charge $\ell = 0$. This procedure was repeated cyclically for $\ell = 0$, $1$, $-1$, and back to $0$ (see Fig. \ref{fig:3}(a)). During the experiment, we observed TL-induced structural changes by tracking the characteristic asymmetrical behavior in diffraction patterns (see Figs. \ref{fig:1}(b) and \ref{fig:3}(a)).

X-rays scattered by a single CBNO nanoflake satisfying the (002) Bragg condition were recorded on an area detector (see Fig.~\ref{fig:3}(a) and {\bf Supplemental Data}, S11). The experimental geometry, combined with the random orientation of the exfoliated nanoflakes, ensured that the (002) Bragg reflections from separate particles were well separated, allowing individual reflections to be isolated on the detector. The nanoflake was kept under continuous coherent TL UV illumination before the imaging experiment. From the coherent diffraction data, we reconstructed both the 3D distribution of the displacement field and the electron density ($\rho(x, y, z)$), along [002], $u_{002}(x,y,z)$, in an individual nanoflake with 14~nm spatial resolution (see {\bf Supplemental Data} Figs. S14, S15, and S16).

Figure~\ref{fig:3}(b) shows a cross-section of the 3D displacement field [$u_{002}(x,y,z = z_0)$] in the FE nanoflake. The [002] direction is approximately along the $z-$axis, while the X-ray beam is almost parallel to the $x-$axis.  
 Depending on the topological polar structure type, the angular distribution will have distinct features associated with a topological invariant known as the topological FE charge winding number:
\begin{equation} \label{windN} 
\mathscr{Q} = \frac{1}{8\pi} \int dA_i \epsilon_{ijk} \frac{\sum_{\kappa \alpha} Z^*_{\kappa, \alpha} u_{\kappa \alpha}(r)}{\left| \sum_{\kappa \alpha} Z^*_{\kappa, \alpha} u_{\kappa \alpha}(r) \right|} \cdot \left[ \partial_j \left( \frac{\sum_{\kappa \beta} Z^*_{\kappa, \beta} u_{\kappa \beta}(r)}{\left| \sum_{\kappa \beta} Z^*_{\kappa, \beta} u_{\kappa \beta}(r) \right|} \right) \times \partial_k \left( \frac{\sum_{\kappa \gamma} Z^*_{\kappa, \gamma} u_{\kappa \gamma}(r)}{\left| \sum_{\kappa \gamma} Z^*_{\kappa, \gamma} u_{\kappa \gamma}(r) \right|} \right) \right].
\end{equation}
This is a quantized value associated with FE defects such as vortices or skyrmions that measure the number of times the displacement vector field (or related order parameter) wraps around a given point or region in space. These topological features can influence the electronic properties and are important for applications in topological quantum computing and advanced electronic devices\cite{yadav2016observation,ma2018controllable}. In Eqn.\ref{windN} $A_i$ is the surface of a small volume containing the polar structure, $\epsilon_{ijk}$ is the Levi-Civita symbol, and $Z^*_{\kappa, i \alpha}$ is the Born effective charge tensor component for atom $\kappa$, relating the $i$-th component of the polarization to the $\alpha$-th component of the displacement of atom $\kappa$, obtained from DFT computations (see \textbf{Materials and Methods} and {\bf Supplemental Data} S9, and Figs. S8, S9, S10, and S11) for the measured displacement field in Fig.~\ref{fig:3}(b). Displacement fields generated by FE vortices are characterized by a winding number $\mathscr{Q} = \frac{1}{2\pi} \oint_C \nabla \theta \cdot d\mathbf{l} = +1$ around the vortex core in 2D (Supplementary Section S12), regardless of their polarity (clockwise or anticlockwise)~\cite{sanchez20242d}. The angle $\theta = \arctan\left(\frac{\delta_{y}u_{002}}{\delta_{x}u_{002}}\right)$ is related to the components of the displacement vector $\mathbf{u_{002}}(r)$. For polar BPs, the winding number captures the change in the displacement (polarization) direction in three dimensions. For anti-polar Bloch points (ABPs), the winding number captures the reversal in the polarization direction. The integral remains the same but yields a negative winding number due to the opposite orientation. Merons and anti-merons are characterized by half-integer winding numbers $\mathscr{Q} = \pm \frac{1}{4\pi} \oint_C \nabla \theta \cdot d\mathbf{l}$ and a discontinuous polarization field.

By inspecting Fig.~\ref{fig:3}, we identified topological polar structures such as vortices, merons, and BPs in the displacement field and electron density maps obtained by BCDI. Figure~\ref{fig:3}(b) shows that at time $t_0$, two types of antiphase domains, $\alpha$ and $\beta$, can form in CBNO due to the presence of two bifurcation options. Combined with two possible in-plane polarization directions~\cite{guo2021unit,fang2022hybrid} ("$+$" being parallel to the $a$ and $b$-axes and "$-$" being antiparallel to them), there are a total of four antiphase-FE $\mathrm{Z}_2 \times \mathrm{Z}_4$ domains ($\alpha+$, $\beta-$, $\alpha-$, and $\beta+$) existing in this system under $\ell = 0$ (see \textbf{Supplemental Figure} S17). 
\begin{figure}[!t]
\centering
\includegraphics[width=0.95\textwidth]{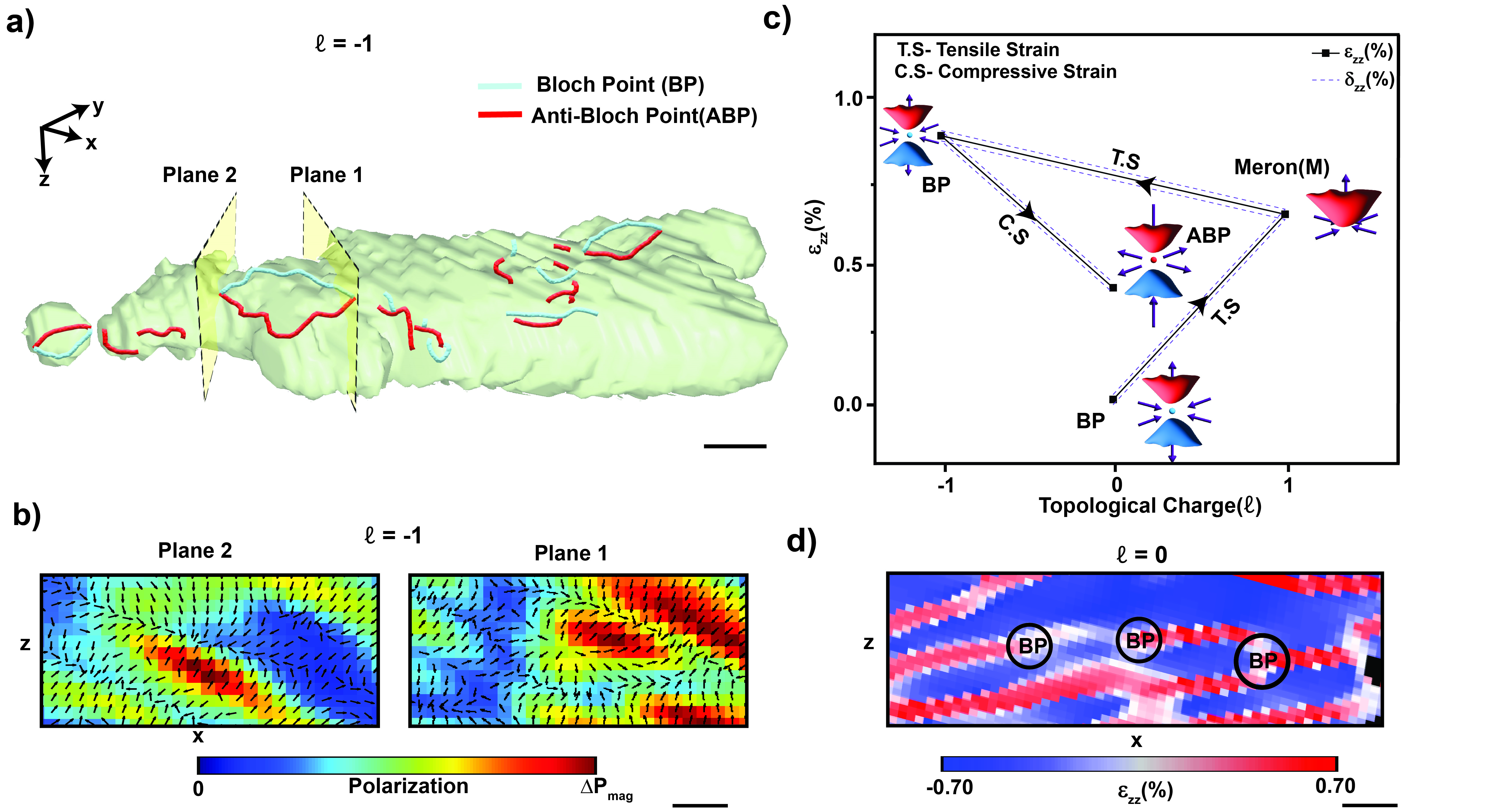}
\caption{\textbf{Twisted light-induced polar Bloch point transitions and strain evolution.} \textbf{(a)} BCDI rendering of the magnitude of the toroidal moment density under OAM of \(\ell = -1\), capturing the topology formed by Bloch and anti-Bloch points. \textbf{(b)} 2D slices depicting the density of BPs and ABPs in the toroidal field, marked by planes in \textbf{(a)}, with the normalized magnitude of polarization $\Delta P_{mag}$vectors overlaid. \textbf{(c)} Hysteresis-like loop capturing OAM field-induced strain \(\epsilon_{zz}\) and BP transition states in CBNO. \textbf{(d)} BCDI strain map under an OAM field of \(\ell = 0\) showing that BPs are associated with regions of reduced polarization magnitude or significant strain gradients. Scale bars correspond to 100 nm.}
\label{fig.4}
\end{figure}

Unlike trimerization observed in other improper FE materials such as the hexamanganites and hexaferrites~\cite{das2014bulk}, where three-fold symmetric domains are formed, the transition in CBNO leads to an in-plane FE domain structure shown in Fig.~\ref{fig:3}.

In Fig.~\ref{fig:3}(b) and Fig.~\ref{fig:3}(d), we qualitatively analyze the polarization vortex by inspection of the displacement field. Approximately 50~nm higher, in-plane Bi-ion displacement due to LG$_\ell$ electric field of $\ell = 0$ induces a periodic strain modulation (Fig.~\ref{fig:3}) in the $xy-$plane above the vortex and creates tensile strain due to its larger lattice constant. However, approximately 50 nm below, the material compensates for this by creating a predominantly compressive strain with a smaller lattice constant. By looking at multiple planes above and below along the $c-$direction, within the spatial resolution of the experiment, we corroborate this periodic modulation of tensile and compressive strain in the $xy-$plane. The strain maps also show intersecting domains that form fork-like patterns in the $yz-$plane. These fork-like features not only guarantee the formation of the vortices but also account for the observed polar BPs (Fig.~\ref{fig:3}(d,e)).

The magnitude of the LG$_\ell$ electric field $E_{OAM} = 10~\text{kV/cm}$ for $\ell = 0$ is smaller than the reported coercive field \cite{fang2022hybrid} for CBNO (see {\bf Supplemental Data}, Section S7, Table S1, Fig. S6), thus explaining why we do not observe significant FE domain switching. To investigate how the observed strain modulation affects the FE polarization, we measured the relative shifts of $\alpha$ and $\beta$ domains (\textbf{Supplemental Data}, Figs. S17 and S18) under the TL induced electric field. We notice that at time $t_{1}$ under an OAM field of $\ell = 1$, the $\alpha^+$ and $\beta^+$ domains shift by vectors of $1/5[110]$ and $-1/5[110]$, respectively, relative to the $\beta^-$. This corresponds to polar displacements as a result of $\text{Bi}^{3+}$ shift in the range of 0.25–0.6 \AA, which is consistent with what is found in bulk CBNO \cite{guo2021unit}.

The evolution of the complex polar textures shown in the planar views of the CBNO displacement field and magnitude (Fig.~\ref{fig:3}) under the vortex electric field $\mathbf{E}_{OAM}$ can be better assessed by quantitative inspection of the three-dimensional BCDI reconstructions. Besides the winding number $\mathscr{Q}$, we can describe the topological structure in terms of a non-zero toroidal moment \cite{wang2020polar, zhang2017nanoscale, luk2020hopfions}, defined as $\mathbf{G} = \frac{1}{2N} \sum_{i=1}^{N} \mathbf{r}_i \times \mathbf{P}_i,$ where \(\mathbf{P}_i\) is the local dipole moment located at \(\mathbf{r}_i\) and \(N\) is the number of dipoles (cells). \(\mathbf{P}_i\) is estimated using the Born effective charge tensor, \(P_i = \sum_{\kappa \alpha} Z^*_{\kappa, i \alpha} u_{\kappa \alpha}\).

We compute the FE toroidal moment from the reconstructed polarization. Regions of the large toroidal moment are plotted in Fig.~\ref{fig.4}(a), where several ‘tubes’ and loops corresponding to the cores of vortices, antivortices, BPs, and ABPs are visible (see {\bf Supplemental Data}, Figs. S23, S24, S25, S26, and S27).
Unlike in incompressible fluids, where the divergence must vanish, a non-zero divergence of the FE polarization due to complex topological structures like FE vortices and BPs signifies the presence of bound charges that play a role in the formation and stability of these structures under the TL-induced vortex field. BPs and ABPs are identified by positive (red) and negative (blue) values within the reconstructed toroidal moment field. We observe a large number of three-dimensional loops (Fig.~\ref{fig.4}(a)) that resemble vortex rings under LG$_\ell$ electric field of $\ell = 1$. 
We consider the case of one such loop, identified by plotting an isosurface corresponding to a maximum threshold of $\pm 0.05P_{s}$ (Fig.~\ref{fig.4}(a)). This loop is located in the vicinity of a single Bloch (anti-Bloch) point, spanning a region of the CBNO crystal. This loop texture represents a departure from the coaxial string-like topology observed under the TL-induced vortex field of $\ell = 0$ and $\ell = 1$ ({\bf Supplemental Data}, Figs. S25, S26, and S27). The 2D cut planes in Fig.~\ref{fig.4}(b) show the vector field of change in $P_{mag}$ (see \textbf{Supplemental Data} Section S16), allowing us to identify topological features such as head-to-head convergence, tail-to-tail divergence, BPs, ABPs, and merons. 

In Ref.~\cite{wang2024polar}, polar BPs were observed in strained PbTiO$_{3}$ films using phase-field simulations and aberration-corrected scanning transmission electron microscopy at the locations where a vortex core intersected a domain wall. Similarly, we find that the BP pair is located at the intersection of the vortex-antivortex loop with a domain wall separating regions of opposite strain magnitude (Fig.~\ref{fig.4}(d)). It was predicted in the same reference that polar BPs can be stabilized using substrate-induced tensile strain in FE films. In Ref.~\cite{sanchez20242d}, a periodic shear strain landscape was generated by stacking freestanding FE perovskite layers with controlled twist angles, akin to Moiré lattices \cite{lau2022reproducibility}. An important remark is that in both cases above, the inter-layer strain in the film cannot be dynamically tuned.

Here, in Fig.~\ref{fig.4}(c), we observe that under LG$_\ell$ electric field transitioning from \(\ell = 0\) to \(\ell = 1\), then \(\ell = -1\), and back to \(\ell = 0\), we can manipulate the global and local strain within the CBNO crystal in a controllable manner ({\bf Supplemental Data}, Figs. S19 and S20). Under a global tensile strain \(\varepsilon_{zz}\) of 0.7\%, there is a transition from BPs to merons. A slight increase in the tensile strain to \(\varepsilon_{zz}\) of 0.75\%, by changing the vortex field to \(\ell = -1\), reverts to a global Bloch point state. Subsequent compressive strain back to \(\ell = 0\) transitions to a global ABP state within the CBNO crystal ({\bf see Supplemental Data}, Tables S2 and S3 and Fig. S22). Figure S20 shows the local mapping of the strain modulations and induced shear strains that mediate the observed BP transitions. 
It is worth noting that such controlled strain manipulation and the observed strain landscape are unique, as they cannot be attained, to the best of our knowledge, either by epitaxial strain or by any pattern of externally applied stresses or electrodes.

In Figs.~\ref{fig:3} and ~\ref{fig.4}, we have presented experimental evidence of the formation of polar topological states in the quasi-2D FE nanoflake, with emerging polarization vortex structures that are axially symmetric at time \(t_{0}\) under LG$_\ell$ electric field of \(\ell = 0\). However, the creation of a vortex costs additional energy due to the existence of the vortex core where the FE is suppressed. Thus, under UV LG$_\ell$ beam illumination of \(\ell = 1\), the singularity at the vortex core can be completely eliminated by the escape of the vector field into the third dimension (see Fig. S19) along the vortex axis, leading to the emergence of chiral meron-like structures in the flake. Under TL illumination of \(\ell = -1\), the system returns to a BP state, while subsequent compressive strain allows the polar structure to transition to an ABP state under a TL-induced LG$_\ell$ field of \(\ell = 1\).

\begin{figure}[!t]
\centering
\includegraphics[width=0.95\textwidth]{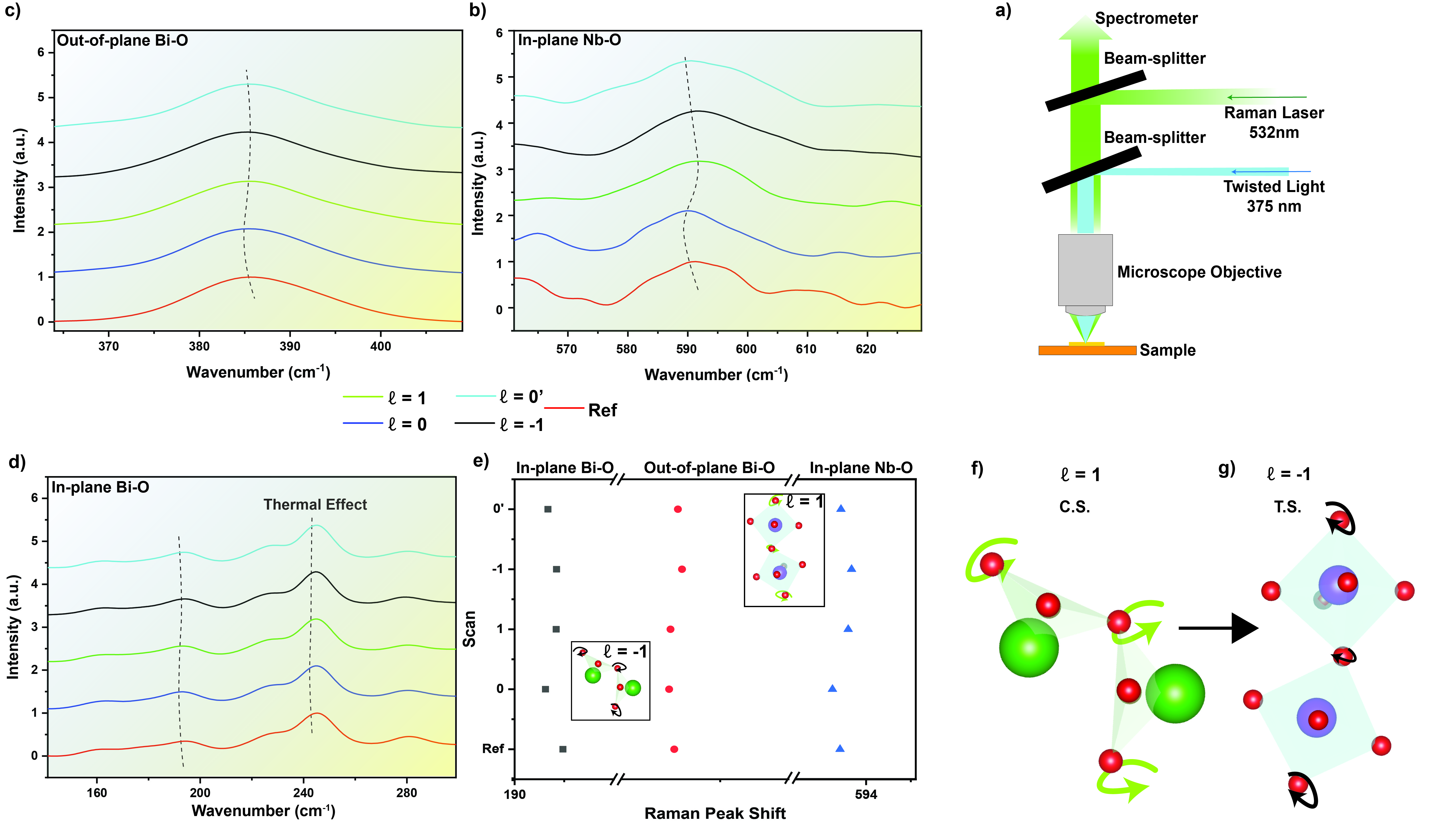}
\caption{\textbf{Twisted-light dependent Raman spectroscopy on CsBiNb$_2$O$_7$:} \textbf{(a)} Schematic of the in-situ Raman experimental setup designed to probe photo-induced structural changes, modulation of phonon modes, and domain dynamics in a CBNO nanoflake under LG$_\ell$ light illumination. The setup utilizes a 532~nm Raman laser with a 375~nm coherent twisted light of varying topological charges. \textbf{(b)} Line plots showing the behavior of the out-of-plane phonon mode for the Bi-O layer. The spectra demonstrate reversible shifts in peak positions, characterized by a tensile shift from the reference to $\ell = 0$, compression under $\ell = 1$, and a reversion to a tensile state under $\ell = -1$, with minimal changes upon returning to $\ell = 0$. This suggests significant OAM-induced modulation of vibrational states. \textbf{(c)} Line plots illustrating the out-of-plane phonon modes of the Bi-O layer, including isolated Bi atom responses. The plots indicate compression effects attributed to thermal influences from continuous illumination, highlighting the sensitivity of Bi phonon modes to UV LG$_\ell$ light and thermal input. \textbf{(d)} Line plots for the in-plane phonon modes of the Bi-O layers, showing reversible Raman peak shifts similar to those observed in the out-of-plane Bi-O modes, underscoring the symmetric impact of OAM on in-plane Bi-O vibrations. \textbf{(e)} Summary of Raman peak shifts across different modes when interacting with twisted light carrying various topological charges. Insets depict the structural and bond modifications along with torsional strain effects induced by twisted light, demonstrating the impact of OAM on lattice dynamics (refer to {\bf Supplemental Data}, Figs. S10, S12, S31, and S32 for detailed analysis). \textbf{(f--g)} Schematic representations of interactions between twisted light with topological charges $\ell = 1$ (Compressive Strain) and $\ell = -1$ (Tensile Strain), with Bi-O tetrahedra and Nb-O octahedra, respectively. These illustrate the torsional and structural changes imparted by non-zero OAM, contributing to the understanding of vortex-light-induced Raman activity.}
\label{fig.5}
\end{figure}

In addition to the observed TL-induced OAM field-induced strain and manipulation of real-space FE (FE) polar structures from BCDI, OAM-dependent Raman measurements on CBNO suggest a possible correlation with twisted UV light-induced excitation of the zone-center and zone-boundary modes in CBNO (see details in \textbf{Materials and Methods} and {\bf Supplemental Data}, Figs. S9, S10 and S12).\\
Figure~\ref{fig.5}(a) illustrates the experimental setup for OAM-dependent Raman measurements on CBNO. Details of the experimental methodology can be found in \textbf{Materials and Methods}. A reference scan was initially performed to confirm the unperturbed crystal structure of CBNO, as shown in \textbf{Supplemental Data} Fig.~S29. Subsequent Raman scans were conducted under various OAM conditions, similar to BCDI measurements (\(\ell = 0\), \(\ell = 1\), \(\ell = -1\), and returning to \(\ell = 0\)). We monitored both the in-plane zone-center phonon mode, reported around 190 cm\(^{-1}\) for the Bi-O bond \cite{fennie2006FE}, and the out-of-plane phonon mode, as depicted in Figs.~\ref{fig.5}(d) and \ref{fig.5}(c), respectively. Illumination with \(\ell = 0\) resulted in tensile strain along the c-axis, while the OAM field with \(\ell = 1\) induced a compressive effect. These findings are consistent with the BCDI measurements. A reversible behavior was observed with \(\ell = -1\), and returning to \(\ell = 0\) resulted in a Raman spectral shift that matched the initial state. This reversibility was also seen in the zone-boundary phonon mode, which has been reported around 580 cm\(^{-1}\) for the Nb-O in-plane Raman peak \cite{jiang2022giant}, as shown in Fig.~\ref{fig.5}(b). The consistent nature of these trends was confirmed by tracking the Raman peak position at full width at half maximum (FWHM), as illustrated in Fig.~\ref{fig.5}(e).

Schematics showing the effects of twisted light (TL)-induced OAM fields (\(\ell = 1\) and \(\ell = -1\)) on the Bi-O and Nb-O bonds are presented in Figs.~\ref{fig.5}(f) and \ref{fig.5}(g), respectively. The schematic is inspired by the data shown in Figs. ~\ref{fig.5}(b),(c) ~\&~ (d), where we observe a reversible tensile and compressive behavior in strain in the phonon modes. This reiterates and emphasizes the ability of the TL-induced electric fields to control the bending and breathing of the phonon modes in the nanocrystal. An inset illustrating the inverse case is shown in Fig.~\ref{fig.5}(e), and a detailed analysis of the shifts in polarization is depicted in {\bf Supplemental Data}, Fig.~S9. The Raman spectroscopy results align with the proposed interaction mechanism between TL and FE polarization in CBNO. Qualitatively, comparisons between BCDI and Raman spectroscopy reveal a hysteresis-like behavior in both strain and Raman peak shifts. This is analogous to the traditional hysteresis observed between polarization and electric field in FEs, where
 the beam topological charge acts similarly to a variable electric field. However, the complex nature of the LG$_\ell$ electric field provide a wider space to manipulate the sample. This demonstrates the potential for controlling FE topologies using TL, opening new avenues for research into vortex-light--matter interactions.

We propose that the interaction mechanism involves the coupling of TL with the electronic structure, leading to shifts in phonon modes within the CBNO nanoflake. One potential interaction mechanism is intra-band transitions, as discussed in \textbf{Supplemental Data}, S8 and S10. We suggest that electron-mediated shifts in the phonon mode structure result in the observed changes in strain, where a preferential shift in the electronic structure is noted (see {\bf Supplemental Data}, Fig.~S13) within the tight-binding approximation of the CBNO band structure. In both BCDI and Raman spectroscopy experimental techniques, the effects are observed post electron-phonon renormalization, leading to the observed Raman peak shifts, FE polarization, and strain behavior.

Our experimental results, supported by theoretical analysis, reveal that the OAM states $\ell = 0$ and $\ell = 0'$ correspond to a uniform electric field distribution, while $\ell = 1$ and $\ell = -1$ result in a non-uniform electric field when illuminating the nanoflake. In both cases, the strength and direction of the electric field vary depending on the beam waist and the topological charge. Under a homogeneous electric field corresponding to $\ell = 0$, a stable two-dimensional vortex structure was observed at time $t = 0$, with the vortex core aligned along one of the crystallographic axes [001] (see Fig.~\ref{fig:3}(c)). A small quasi-static field applied at $t = 0$ stabilized an axial vortex core, which is topologically protected by two antiphase nanodomains with nearly homogeneous spontaneous polarization (see {\bf Supplemental Data} Section S8 and Fig. S21).

Similar to ferromagnetic vortices, the core of the observed ferroelectric vortex behaves like a dipole, offering the potential to rotate the vortex axis under the influence of a small quasi-static electric field, whether homogeneous or inhomogeneous. Due to strong ferroelectric anisotropy, which hinders the rotation of the dipolar vortex core and tends to align it along one of the crystallographic directions, the orientation of the core in the field is determined by the minimization of dipolar and anisotropy energies. As the vortex core is coupled to the antiphase domains, the plane of the vortex, which is perpendicular to the core axis, can rotate along with the core. The core size is sensitive to variations in the TL-induced electric field, and as these variations occur, the nanoflake gradually transitions toward a single-domain state, as observed at time $t = t_1$.

Manipulating FE polarization in quasi-2D materials like CsBiNb$_2$O$_7$ using incident vortex fields generated by coherent twisted ultraviolet light offers distinct advantages over using pulsed terahertz infrared photons, as traditionally employed in 3D perovskite materials like SrTiO$_3$. Twisted ultraviolet light carries topological charge, imparting angular momentum to the electrons and ions in the material, leading to precise interactions with ferroelectric polarization. This helical phase can stabilize transitions between Bloch points and merons by aligning polarization vectors and modifying the energy landscape, thereby lowering the energy barriers for these transitions. The higher spatial precision and photon energy of UV light allow for more effective and localized polarization changes compared to terahertz infrared pulses. Additionally, vortex fields carrying orbital angular momentum can selectively interact with specific phonon modes or electronic states that are not easily accessible with terahertz infrared pulses, providing a unique tool for manipulating ferroelectric polarization.

In 3D perovskite oxides systems ~\cite{basini2022terahertz,li2019terahertz,gao2024effective}, pulsed terahertz infrared photons can resonantly excite specific phonon modes, inducing transient changes in ferroelectric polarization and numerous emergent phenomena~\cite{liu2012terahertz,fausti2011light,basini2024terahertz}. However, the additional degree of freedom and increased pathways for energy dissipation in 3D systems make it more challenging to achieve the same level of control and precision as in quasi-2D systems. Therefore, using electric field of coherent twisted ultraviolet light to manipulate ferroelectric polarization in quasi-2D CsBiNb$_2$O$_7$ provides superior control and effectiveness, enabling the stabilization of complex topological transitions.
\section*{CONCLUSION AND OUTLOOK}

We have demonstrated that the orbital angular momentum electric field from twisted light can dynamically induce strains (both tensile and compressive) up to $\sim$ 0.8\% along the c-axis in individual quasi-2D ferroelectric CsBiNb$_2$O$_7$ crystals. These strains are quantifiable, with nanometer resolution, and can be controlled reversibly, creating significant strain gradients. This phenomenon is potentially universal to quasi-2D ferroelectric systems with weak out-of-plane bonding along the stacking direction. For example, in CsBiNb$_2$O$_7$, the reported piezoelectric coefficient $d_{33}$ (8 pC N$^{-1}$)\cite{chen2015FE} is significantly smaller than in typical 3D perovskite oxides (e.g., PbTiO$_3$, BaTiO$_3$\cite{shi2023enhanced}, BiFeO$_3$). Our observations show that topologically protected polar structures and transitions from Bloch points to merons occur under a light-carrying orbital angular momentum electric field, mediated by induced tensile strains. These findings align with earlier reports of polar Bloch points in strained epitaxial thin films. However, in epitaxial thin films, the mechanical boundary conditions are rigid and determined by the atom-on-atom replication of the substrate structure, leaving little room for modification. Additionally, reversible and dynamic control of torsional, shear, or inhomogeneous strain patterns is not easily achievable via epitaxy.

In summary, we have found that it is possible to induce non-trivial ferroelectric textures and polar structures, as well as strains in quasi-2D freestanding ferroelectric crystalline flakes. The driving force is the coupling between the twisted light’s spatially varying electric field and the crystal, which effectively imposes electromechanical (stress and vectorial electric field) boundary conditions. These couplings induce large strains and strain gradients in the ferroelectric material, resulting in vortex-like and Bloch-type modulations of the homogeneous polarization state. Furthermore, the nature of these 2D and 3D topological structures can be largely tuned by controlling the twisted light's topological charge.

Moreover, the interaction of twisted light with ferroelectrics potentially opens new avenues for research into the dynamical twist-multiferroics \cite{basini2024terahertz} and the braiding of polar 1D structures. Upgrades in the coherence~\cite{kerby2023advanced, fohtung2018bragg} of synchrotrons and the development of time-resolved orbital angular momentum pump orbital angular momentum-probe Bragg coherent diffractive imaging techniques that require twisted X-rays~\cite{nazirkar2024coherent} could potentially provide deeper insights into the dynamical control of novel states of matter that are unattainable by conventional thermodynamic methods. Additionally, the development of new Bragg coherent diffractive imaging methods to access in-plane strain and orbital angular momentum-dependent Raman spectroscopy to capture chiral or more nuanced twisted-light interactions could represent a paradigm shift in engineering beyond 2D. This includes the potential for dynamical manipulation of the twist angle in which other modes of twisted light, such as higher-order Bessel beams with higher penetration, could play important roles. Shedding further light on these advances would be a great challenge, for both theory and experiment, but is beyond the scope of the present work.


\section*{MATERIALS AND METHODS} \label{MatandMethods}

\subsection*{Fabrication of freestanding quazi-2D flakes}

Free-standing CBNO 2D flakes were synthesized using a molten-salt assisted method~\cite{guo2021unit, jiang2022giant}. Further, pre-characterization was carried out to confirm the stoichiometry using XPS ({\bf Supplemental Data}, S4), morphology using XRD and SEM ({\bf Supplemental Data}, S2 and S3) and bandgap using transmission spectroscopy ({\bf Supplemental Data}, S5). 

\subsection*{Twisted light Bragg coherent diffractive imaging experiments}

The Si (111) monochromator at the Advanced Photon Source's sector 34-ID-C was utilized to isolate coherent X-ray photons with an energy of 9.0 keV. The beam's monochromaticity was defined by an energy bandwidth of 1~eV, and it exhibited a 0.7~$\mu$m transverse coherence length. The X-ray beam was focused onto the sample using a pair of Kirkpatrick-Baez mirrors situated after the beam-defining aperture. For this experiment, the beam size was set to 700 nm by 700 nm. The principle of the Bragg coherent diffraction experiment is illustrated in Fig.~S3. A Medipix2 CMOS X-ray detector was positioned around the diffraction sphere using a motorized arm. The detector was aligned with the outgoing (002) characteristic Bragg reflection from the CBNO sample. To magnify the interference fringes in the diffraction pattern, the detector was placed 1.2 m away from the sample. An evacuated flight tube positioned in the path from the sample to the detector helped minimize photon scattering losses in the air. For such a photon-starving technique as nanoscale Bragg coherent diffractive imaging, the use of an evacuated flight tube, high sensor gain, and the detector's photon counting mode are essential. An Ultra-Violet laser of 375nm, guided via a confocal system available at the 34-IDC beamline, was focused down to 10~$\mu$m at a normal incidence to the isolated CBNO flake of interest. During the collection of rocking curves, the sample was continuously illuminated. We employed a spatial light modulator (SLM) to select the optical beam's topological charge, thereby controlling the OAM and torque transferred to the crystal (see {\bf Supplemental Data}, S6 and Fig. S3). Rocking curves were collected as a series of 2D diffraction patterns near the CBNO  ${002}$ Bragg peak, corresponding to 2$\theta$ = 13.26\degree, with a scanning range of $\Delta\theta = \pm 0.2\degree$ in the vicinity of the Bragg peak. A total of 360 patterns were gathered in a single rocking curve. The dataset for the virgin state (topological charge, $\ell = 0$) was obtained before cycling the CBNO crystal. The subsequent states under continuous LG$_\ell$ beam illumination, $\ell = 1$, $\ell = -1$ and $\ell = 0$ were recorded after 30 cycles of $\ell = 0$ illumination and release. This takes about 30 mins for one single 3D scan to be collected. Thus for all OAM applications $T = t_0=t_1=t_2=t_3=30$~mins. This ensured that the system was entirely in equilibrium under laser illumination.

\subsection*{Phase retrieval process}
The collected 3D diffraction patterns were inverted from reciprocal into real space using iterative phase retrieval algorithms\cite{shi2023enhanced}. This gives us the complex wavefield post diffraction, which is given by $\rho(\textbf{r}) = \rho_0 (\textbf{r})e^{(i\phi(\textbf{r}))}$ where $\phi(\textbf{r})$ is the phase and $\rho(r)$ is the amplitude of the wave. The phase carries the information of about the displacement field $u_{002}$, which can be determined by using $\phi(\textbf{r})$ = \textbf{G$_{002}$}.\textbf{u$_{002}$}(\textbf{r}). The shape and size of the nanocrystal are estimated from the reconstructed Bragg Electronic Density($\rho$(\textbf{r})). For a given applied TL topological charge, real-space images of the CBNO nanocrystal were reconstructed with approximately 33 nm spatial resolution as determined by the phase retrieval transfer function (PRTF) (see {\bf Supplemental Data}, Fig. S28). The 3D nature of the reconstructed nanocrystal allows us to slice through the particle and analyze the different FE topologies. Fienup's hybrid input-output (HIO) approach was the basis for iterative phase retrieval techniques, which were further enhanced by randomized overrelaxation. When the measurement locations are close enough to one another to satisfy the oversampling criterion, an important stage in the process is to reverse the diffraction data using a computer algorithm. The first stage is to assume a three-dimensional support volume where the sample density will be completely limited. These techniques impose a Fourier transform, both forward and backward, between the real and reciprocal spaces, imposing an intensity mask constraint in the former and a support constraint in the latter. 

\subsection*{Twisted-light Raman spectroscopy}
A 532 nm wavelength light from the Witec Alpha 300R series was used to perform the Raman Measurements. The Witec Alpha 300R was modified with a free beam coupler to introduce the twisted light of wavelength 375~nm on to the sample surface. During the collection of a single spectrum, the sample was continuously illuminated. We employed a spatial light modulator (SLM) to select the optical beam's topological charge, thereby controlling the OAM and torque transferred to the crystal (see {\bf Supplemental Data}, S18 and Fig.~S29). An Andor CCD was used to collect and count photons scattered from the CBNO nanoflake. A spectroscope grating of 1800 gr/mm was used to obtain a high-resolution spectrum of the sample under various topological charges. A Zeiss 100x microscope objective was used for focusing the laser beams onto the nanoflake as shown in Fig.~S29. The integration time of 1~$s$ was used to collect 100 accumulations of the spectrum to obtain a clean spectrum with a low signal-to-noise ratio. The reference spectrum was collected before cycling the system. The subsequent spectrums were collected under continuous illuminations of $\ell = 1$, $\ell = -1$ and $\ell = 0$ were collected. 

\subsection*{ First-principles calculations}

We performed first-principles calculations using the CASTEP (Cambridge Serial Total Energy Package) \cite{segall2002first} and Quantum ESPRESSO (QE) \cite{giannozzi2009quantum} packages. The calculations involving the application of an external light beam carrying Orbital Angular Momentum (OAM) with different topological charges ($\ell$) were performed using CASTEP. The experimental electric field component of the OAM beam was explicitly included in these calculations. We employed the Perdew-Burke-Ernzerhof (PBEsol) exchange-correlation functional within the generalized gradient approximation (GGA) \cite{perdew1996generalized} for all calculations. We used norm-conserving pseudopotentials and a plane-wave basis set with a kinetic energy cutoff of 600 eV. The pseudopotentials for Cs, Bi, Nb, and O were taken from the CASTEP library.\\
The Brillouin zone was sampled using a $6 \times 6 \times 6$ Monkhorst-Pack grid \cite{monkhorst1976special}. The self-consistent field (SCF) calculations were converged to an energy threshold of $1 \times 10^{-6}$ eV. The initial molecular structure of CsBiNb$_2$O$_7$ was obtained from experimental theoretical data predictions available on the Materials Project. The structures were optimized using density functional theory (DFT) as implemented in Quantum ESPRESSO. The Perdew-Burke-Ernzerhof (PBEsol) exchange-correlation function was chosen to describe the electron-electron interactions. The unit cell parameters are: 
$a = 11.87380$~\AA, $b = 5.45730$~\AA, $c = 5.55750$~\AA, with angles $\alpha = 90^\circ$, $\beta = 90^\circ$, $\gamma = 90^\circ$. 
Atomic positions were fully relaxed until the forces on each atom were less than 0.01~eV/\AA, and the stress tensor was below 1$\times10^{-3}$~GPa.
The polarizability was computed using density functional perturbation theory (DFPT) \cite{baroni2001phonons}. Infrared and Raman spectra were obtained by calculating the dynamical charges and the Raman tensors, respectively. The dielectric function and the Raman activities were computed to simulate the IR and Raman spectra.

\bibliography{sample}

\begin{thebibliography}{10}
\urlstyle{rm}
\expandafter\ifx\csname url\endcsname\relax
  \def\url#1{\texttt{#1}}\fi
\expandafter\ifx\csname urlprefix\endcsname\relax\def\urlprefix{URL }\fi
\expandafter\ifx\csname doiprefix\endcsname\relax\def\doiprefix{DOI: }\fi
\providecommand{\bibinfo}[2]{#2}
\providecommand{\eprint}[2][]{\url{#2}}

\bibitem{von2018probing}
\bibinfo{author}{von Hoegen, A.}, \bibinfo{author}{Mankowsky, R.},
  \bibinfo{author}{Fechner, M.}, \bibinfo{author}{F{\"o}rst, M.} \&
  \bibinfo{author}{Cavalleri, A.}
\newblock \bibinfo{journal}{\bibinfo{title}{Probing the interatomic potential
  of solids with strong-field nonlinear phononics}}.
\newblock {\emph{\JournalTitle{Nature}}} \textbf{\bibinfo{volume}{555}},
  \bibinfo{pages}{79--82} (\bibinfo{year}{2018}).

\bibitem{liu2012terahertz}
\bibinfo{author}{Liu, M.} \emph{et~al.}
\newblock \bibinfo{journal}{\bibinfo{title}{Terahertz-field-induced
  insulator-to-metal transition in vanadium dioxide metamaterial}}.
\newblock {\emph{\JournalTitle{Nature}}} \textbf{\bibinfo{volume}{487}},
  \bibinfo{pages}{345--348} (\bibinfo{year}{2012}).

\bibitem{fausti2011light}
\bibinfo{author}{Tobey, R.}, \bibinfo{author}{Dienst, A.},
  \bibinfo{author}{Pyon, S.}, \bibinfo{author}{Takayama, T.} \&
  \bibinfo{author}{Takagi, H.}
\newblock \bibinfo{journal}{\bibinfo{title}{Light-induced superconductivity in
  a stripe-ordered cuprate}}.
\newblock {\emph{\JournalTitle{science}}} \textbf{\bibinfo{volume}{331}},
  \bibinfo{pages}{189--191} (\bibinfo{year}{2011}).

\bibitem{li2019terahertz}
\bibinfo{author}{Li, X.} \emph{et~al.}
\newblock \bibinfo{journal}{\bibinfo{title}{Terahertz field--induced
  ferroelectricity in quantum paraelectric srtio3}}.
\newblock {\emph{\JournalTitle{Science}}} \textbf{\bibinfo{volume}{364}},
  \bibinfo{pages}{1079--1082} (\bibinfo{year}{2019}).

\bibitem{basini2022terahertz}
\bibinfo{author}{Pancaldi, M.} \emph{et~al.}
\newblock \bibinfo{journal}{\bibinfo{title}{Terahertz electric-field driven
  dynamical multiferroicity in srtio $ \_3$}}.
\newblock {\emph{\JournalTitle{arXiv preprint arXiv:2210.01690}}}
  (\bibinfo{year}{2022}).

\bibitem{gao2024dynamical}
\bibinfo{author}{Gao, L.}, \bibinfo{author}{Prokhorenko, S.},
  \bibinfo{author}{Nahas, Y.} \& \bibinfo{author}{Bellaiche, L.}
\newblock \bibinfo{journal}{\bibinfo{title}{Dynamical control of topology in
  polar skyrmions via twisted light}}.
\newblock {\emph{\JournalTitle{Physical Review Letters}}}
  \textbf{\bibinfo{volume}{132}}, \bibinfo{pages}{026902}
  (\bibinfo{year}{2024}).

\bibitem{gao2024effective}
\bibinfo{author}{Gao, L.}, \bibinfo{author}{Prokhorenko, S.},
  \bibinfo{author}{Nahas, Y.} \& \bibinfo{author}{Bellaiche, L.}
\newblock \bibinfo{journal}{\bibinfo{title}{Effective gyration of polar vortex
  arrays controlled by high orbital angular momentum of light}}.
\newblock {\emph{\JournalTitle{Physical Review B}}}
  \textbf{\bibinfo{volume}{109}}, \bibinfo{pages}{L121110}
  (\bibinfo{year}{2024}).

\bibitem{kittel1949physical}
\bibinfo{author}{Kittel, C.}
\newblock \bibinfo{journal}{\bibinfo{title}{Physical theory of ferromagnetic
  domains}}.
\newblock {\emph{\JournalTitle{Reviews of modern Physics}}}
  \textbf{\bibinfo{volume}{21}}, \bibinfo{pages}{541} (\bibinfo{year}{1949}).

\bibitem{kittel1946theory}
\bibinfo{author}{Kittel, C.}
\newblock \bibinfo{journal}{\bibinfo{title}{Theory of the structure of
  ferromagnetic domains in films and small particles}}.
\newblock {\emph{\JournalTitle{Physical Review}}}
  \textbf{\bibinfo{volume}{70}}, \bibinfo{pages}{965} (\bibinfo{year}{1946}).

\bibitem{tang2015observation}
\bibinfo{author}{Tang, Y.} \emph{et~al.}
\newblock \bibinfo{journal}{\bibinfo{title}{Observation of a periodic array of
  flux-closure quadrants in strained ferroelectric pbtio3 films}}.
\newblock {\emph{\JournalTitle{Science}}} \textbf{\bibinfo{volume}{348}},
  \bibinfo{pages}{547--551} (\bibinfo{year}{2015}).

\bibitem{ma2018controllable}
\bibinfo{author}{Ma, J.} \emph{et~al.}
\newblock \bibinfo{journal}{\bibinfo{title}{Controllable conductive readout in
  self-assembled, topologically confined ferroelectric domain walls}}.
\newblock {\emph{\JournalTitle{Nature nanotechnology}}}
  \textbf{\bibinfo{volume}{13}}, \bibinfo{pages}{947--952}
  (\bibinfo{year}{2018}).

\bibitem{karpov2017three}
\bibinfo{author}{Karpov, D.} \emph{et~al.}
\newblock \bibinfo{journal}{\bibinfo{title}{Three-dimensional imaging of vortex
  structure in a ferroelectric nanoparticle driven by an electric field}}.
\newblock {\emph{\JournalTitle{Nature Communications}}}
  \textbf{\bibinfo{volume}{8}} (\bibinfo{year}{2017}).

\bibitem{yadav2016observation}
\bibinfo{author}{Yadav, A.} \emph{et~al.}
\newblock \bibinfo{journal}{\bibinfo{title}{Observation of polar vortices in
  oxide superlattices}}.
\newblock {\emph{\JournalTitle{Nature}}} \textbf{\bibinfo{volume}{530}},
  \bibinfo{pages}{198--201} (\bibinfo{year}{2016}).

\bibitem{das2019observation}
\bibinfo{author}{Das, S.} \emph{et~al.}
\newblock \bibinfo{journal}{\bibinfo{title}{Observation of room-temperature
  polar skyrmions}}.
\newblock {\emph{\JournalTitle{Nature}}} \textbf{\bibinfo{volume}{568}},
  \bibinfo{pages}{368--372} (\bibinfo{year}{2019}).

\bibitem{wang2020polar}
\bibinfo{author}{Wang, Y.} \emph{et~al.}
\newblock \bibinfo{journal}{\bibinfo{title}{Polar meron lattice in strained
  oxide ferroelectrics}}.
\newblock {\emph{\JournalTitle{Nature Materials}}}
  \textbf{\bibinfo{volume}{19}}, \bibinfo{pages}{881--886}
  (\bibinfo{year}{2020}).

\bibitem{zhang2017nanoscale}
\bibinfo{author}{Zhang, Q.} \emph{et~al.}
\newblock \bibinfo{journal}{\bibinfo{title}{Nanoscale bubble domains and
  topological transitions in ultrathin ferroelectric films}}.
\newblock {\emph{\JournalTitle{Advanced Materials}}}
  \textbf{\bibinfo{volume}{29}}, \bibinfo{pages}{1702375}
  (\bibinfo{year}{2017}).

\bibitem{luk2020hopfions}
\bibinfo{author}{Luk’Yanchuk, I.}, \bibinfo{author}{Tikhonov, Y.},
  \bibinfo{author}{Razumnaya, A.} \& \bibinfo{author}{Vinokur, V.}
\newblock \bibinfo{journal}{\bibinfo{title}{Hopfions emerge in
  ferroelectrics}}.
\newblock {\emph{\JournalTitle{Nature Communications}}}
  \textbf{\bibinfo{volume}{11}}, \bibinfo{pages}{2433} (\bibinfo{year}{2020}).

\bibitem{salje2014FE}
\bibinfo{author}{Salje, E.} \& \bibinfo{author}{Scott, J.}
\newblock \bibinfo{journal}{\bibinfo{title}{Ferroelectric bloch-line switching:
  A paradigm for memory devices?}}
\newblock {\emph{\JournalTitle{Applied Physics Letters}}}
  \textbf{\bibinfo{volume}{105}} (\bibinfo{year}{2014}).

\bibitem{morozovska2020electric}
\bibinfo{author}{Morozovska, A.~N.} \emph{et~al.}
\newblock \bibinfo{journal}{\bibinfo{title}{Electric field control of
  three-dimensional vortex states in core-shell ferroelectric nanoparticles}}.
\newblock {\emph{\JournalTitle{Acta Materialia}}}
  \textbf{\bibinfo{volume}{200}}, \bibinfo{pages}{256--273}
  (\bibinfo{year}{2020}).

\bibitem{karpov2019nanoscale}
\bibinfo{author}{Karpov, D.} \emph{et~al.}
\newblock \bibinfo{journal}{\bibinfo{title}{Nanoscale topological defects and
  improper ferroelectric domains in multiferroic barium hexaferrite
  nanocrystals}}.
\newblock {\emph{\JournalTitle{Physical Review B}}}
  \textbf{\bibinfo{volume}{100}}, \bibinfo{pages}{054432}
  (\bibinfo{year}{2019}).

\bibitem{shi2023enhanced}
\bibinfo{author}{Shi, X.} \emph{et~al.}
\newblock \bibinfo{journal}{\bibinfo{title}{Enhanced piezoelectric response at
  nanoscale vortex structures in ferroelectrics}}.
\newblock {\emph{\JournalTitle{arXiv preprint arXiv:2305.13096}}}
  (\bibinfo{year}{2023}).

\bibitem{behera2022electric}
\bibinfo{author}{Behera, P.} \emph{et~al.}
\newblock \bibinfo{journal}{\bibinfo{title}{Electric field control of
  chirality}}.
\newblock {\emph{\JournalTitle{Science advances}}}
  \textbf{\bibinfo{volume}{8}}, \bibinfo{pages}{eabj8030}
  (\bibinfo{year}{2022}).

\bibitem{shafer2018emergent}
\bibinfo{author}{Shafer, P.} \emph{et~al.}
\newblock \bibinfo{journal}{\bibinfo{title}{Emergent chirality in the electric
  polarization texture of titanate superlattices}}.
\newblock {\emph{\JournalTitle{Proceedings of the National Academy of
  Sciences}}} \textbf{\bibinfo{volume}{115}}, \bibinfo{pages}{915--920}
  (\bibinfo{year}{2018}).

\bibitem{lovesey2018resonant}
\bibinfo{author}{Lovesey, S.~W.} \& \bibinfo{author}{van~der Laan, G.}
\newblock \bibinfo{journal}{\bibinfo{title}{Resonant x-ray diffraction from
  chiral electric-polarization structures}}.
\newblock {\emph{\JournalTitle{Physical Review B}}}
  \textbf{\bibinfo{volume}{98}}, \bibinfo{pages}{155410}
  (\bibinfo{year}{2018}).

\bibitem{chauleau2020electric}
\bibinfo{author}{Chauleau, J.-Y.} \emph{et~al.}
\newblock \bibinfo{journal}{\bibinfo{title}{Electric and antiferromagnetic
  chiral textures at multiferroic domain walls}}.
\newblock {\emph{\JournalTitle{Nature materials}}}
  \textbf{\bibinfo{volume}{19}}, \bibinfo{pages}{386--390}
  (\bibinfo{year}{2020}).

\bibitem{wang2024polar}
\bibinfo{author}{Wang, Y.-J.} \emph{et~al.}
\newblock \bibinfo{journal}{\bibinfo{title}{Polar bloch points in strained
  ferroelectric films}}.
\newblock {\emph{\JournalTitle{Nature Communications}}}
  \textbf{\bibinfo{volume}{15}}, \bibinfo{pages}{3949} (\bibinfo{year}{2024}).

\bibitem{Jeong2024}
\bibinfo{author}{Jeong, L. J. e. a. Y.~Y., Chaehwa}.
\newblock \bibinfo{journal}{\bibinfo{title}{Revealing the three-dimensional
  arrangement of polar topology in nanoparticles}}.
\newblock {\emph{\JournalTitle{Nature Communications}}}
  \doiprefix\url{10.1038/s41467-024-48082-x} (\bibinfo{year}{2024}).

\bibitem{allen1992orbital}
\bibinfo{author}{Allen, L.}, \bibinfo{author}{Beijersbergen, M.~W.},
  \bibinfo{author}{Spreeuw, R.} \& \bibinfo{author}{Woerdman, J.}
\newblock \bibinfo{journal}{\bibinfo{title}{Orbital angular momentum of light
  and the transformation of laguerre-gaussian laser modes}}.
\newblock {\emph{\JournalTitle{Physical review A}}}
  \textbf{\bibinfo{volume}{45}}, \bibinfo{pages}{8185} (\bibinfo{year}{1992}).

\bibitem{fang2022hybrid}
\bibinfo{author}{Fang, X.}, \bibinfo{author}{Hu, R.}, \bibinfo{author}{Wang,
  Y.}, \bibinfo{author}{Huang, F.-T.} \& \bibinfo{author}{Cheong, S.-W.}
\newblock \bibinfo{journal}{\bibinfo{title}{Hybrid improper ferroelectricity in
  highly cleavable single crystals of dion-jacobson-compound csbinb2o7}}.
\newblock {\emph{\JournalTitle{Advanced Electronic Materials}}}
  \textbf{\bibinfo{volume}{8}}, \bibinfo{pages}{2100929}
  (\bibinfo{year}{2022}).

\bibitem{fennie2006FE}
\bibinfo{author}{Fennie, C.~J.} \& \bibinfo{author}{Rabe, K.~M.}
\newblock \bibinfo{journal}{\bibinfo{title}{Ferroelectricity in the
  dion-jacobson csbinb2o7 from first principles}}.
\newblock {\emph{\JournalTitle{Applied physics letters}}}
  \textbf{\bibinfo{volume}{88}} (\bibinfo{year}{2006}).

\bibitem{chen2015FE}
\bibinfo{author}{Chen, C.} \emph{et~al.}
\newblock \bibinfo{journal}{\bibinfo{title}{Ferroelectricity in dion--jacobson
  abinb 2 o 7 (a= rb, cs) compounds}}.
\newblock {\emph{\JournalTitle{Journal of Materials Chemistry C}}}
  \textbf{\bibinfo{volume}{3}}, \bibinfo{pages}{19--22} (\bibinfo{year}{2015}).

\bibitem{goff2009leakage}
\bibinfo{author}{Goff, R.~J.} \emph{et~al.}
\newblock \bibinfo{journal}{\bibinfo{title}{Leakage and proton conductivity in
  the predicted ferroelectric csbinb2o7}}.
\newblock {\emph{\JournalTitle{Chemistry of Materials}}}
  \textbf{\bibinfo{volume}{21}}, \bibinfo{pages}{1296--1302}
  (\bibinfo{year}{2009}).

\bibitem{snedden2003structural}
\bibinfo{author}{Snedden, A.}, \bibinfo{author}{Knight, K.~S.} \&
  \bibinfo{author}{Lightfoot, P.}
\newblock \bibinfo{journal}{\bibinfo{title}{Structural distortions in the
  layered perovskites csanb2o7 (a= nd, bi)}}.
\newblock {\emph{\JournalTitle{Journal of Solid State Chemistry}}}
  \textbf{\bibinfo{volume}{173}}, \bibinfo{pages}{309--313}
  (\bibinfo{year}{2003}).

\bibitem{hervoches2001two}
\bibinfo{author}{Hervoches, C.}, \bibinfo{author}{Irvine, J. T.~S.} \&
  \bibinfo{author}{Lightfoot, P.}
\newblock \bibinfo{journal}{\bibinfo{title}{Two high-temperature paraelectric
  phases in sr 0.85 bi 2.1 ta 2 o 9}}.
\newblock {\emph{\JournalTitle{Physical Review B}}}
  \textbf{\bibinfo{volume}{64}}, \bibinfo{pages}{100102}
  (\bibinfo{year}{2001}).

\bibitem{jiang2022giant}
\bibinfo{author}{Jiang, J.} \emph{et~al.}
\newblock \bibinfo{journal}{\bibinfo{title}{Giant pyroelectricity in
  nanomembranes}}.
\newblock {\emph{\JournalTitle{Nature}}} \textbf{\bibinfo{volume}{607}},
  \bibinfo{pages}{480--485} (\bibinfo{year}{2022}).

\bibitem{guo2021unit}
\bibinfo{author}{Guo, Y.} \emph{et~al.}
\newblock \bibinfo{journal}{\bibinfo{title}{Unit-cell-thick domain in
  free-standing quasi-two-dimensional ferroelectric material}}.
\newblock {\emph{\JournalTitle{Physical Review Materials}}}
  \textbf{\bibinfo{volume}{5}}, \bibinfo{pages}{044403} (\bibinfo{year}{2021}).

\bibitem{robinson2009coherent}
\bibinfo{author}{Robinson, I.} \& \bibinfo{author}{Harder, R.}
\newblock \bibinfo{journal}{\bibinfo{title}{Coherent x-ray diffraction imaging
  of strain at the nanoscale}}.
\newblock {\emph{\JournalTitle{Nature materials}}}
  \textbf{\bibinfo{volume}{8}}, \bibinfo{pages}{291} (\bibinfo{year}{2009}).

\bibitem{Goff2009}
\bibinfo{author}{Goff, R.~J.} \emph{et~al.}
\newblock \bibinfo{journal}{\bibinfo{title}{Leakage and proton conductivity in
  the predicted ferroelectric csbinb2o7}}.
\newblock {\emph{\JournalTitle{Chemistry of Materials}}}
  \textbf{\bibinfo{volume}{21}}, \bibinfo{pages}{1296--1302},
  \doiprefix\url{10.1021/cm8030895} (\bibinfo{year}{2009}).
\newblock \eprint{https://doi.org/10.1021/cm8030895}.

\bibitem{forbes2017controlling}
\bibinfo{author}{Forbes, A.}
\newblock \bibinfo{journal}{\bibinfo{title}{Controlling light’s helicity at
  the source: orbital angular momentum states from lasers}}.
\newblock {\emph{\JournalTitle{Philosophical Transactions of the Royal Society
  A: Mathematical, Physical and Engineering Sciences}}}
  \textbf{\bibinfo{volume}{375}}, \bibinfo{pages}{20150436}
  (\bibinfo{year}{2017}).

\bibitem{wang2023free}
\bibinfo{author}{Wang, G.} \emph{et~al.}
\newblock \bibinfo{journal}{\bibinfo{title}{Free-space creation of a perfect
  vortex beam with fractional topological charge}}.
\newblock {\emph{\JournalTitle{Optics Express}}} \textbf{\bibinfo{volume}{31}},
  \bibinfo{pages}{5757--5766} (\bibinfo{year}{2023}).

\bibitem{erhard2018twisted}
\bibinfo{author}{Erhard, M.}, \bibinfo{author}{Fickler, R.},
  \bibinfo{author}{Krenn, M.} \& \bibinfo{author}{Zeilinger, A.}
\newblock \bibinfo{journal}{\bibinfo{title}{Twisted photons: new quantum
  perspectives in high dimensions}}.
\newblock {\emph{\JournalTitle{Light: Science \& Applications}}}
  \textbf{\bibinfo{volume}{7}}, \bibinfo{pages}{17146--17146}
  (\bibinfo{year}{2018}).

\bibitem{shen2019optical}
\bibinfo{author}{Shen, Y.} \emph{et~al.}
\newblock \bibinfo{journal}{\bibinfo{title}{Optical vortices 30 years on: Oam
  manipulation from topological charge to multiple singularities}}.
\newblock {\emph{\JournalTitle{Light: Science \& Applications}}}
  \textbf{\bibinfo{volume}{8}}, \bibinfo{pages}{90} (\bibinfo{year}{2019}).

\bibitem{tran2023exploration}
\bibinfo{author}{Tran, V.} \emph{et~al.}
\newblock \bibinfo{journal}{\bibinfo{title}{On the exploration of structured
  light transmission through a multimode fiber in a reference-less system}}.
\newblock {\emph{\JournalTitle{APL Photonics}}} \textbf{\bibinfo{volume}{8}}
  (\bibinfo{year}{2023}).

\bibitem{sanchez20242d}
\bibinfo{author}{S{\'a}nchez-Santolino, G.} \emph{et~al.}
\newblock \bibinfo{journal}{\bibinfo{title}{A 2d ferroelectric vortex pattern
  in twisted batio3 freestanding layers}}.
\newblock {\emph{\JournalTitle{Nature}}} \textbf{\bibinfo{volume}{626}},
  \bibinfo{pages}{529--534} (\bibinfo{year}{2024}).

\bibitem{das2014bulk}
\bibinfo{author}{Das, H.}, \bibinfo{author}{Wysocki, A.~L.},
  \bibinfo{author}{Geng, Y.}, \bibinfo{author}{Wu, W.} \&
  \bibinfo{author}{Fennie, C.~J.}
\newblock \bibinfo{journal}{\bibinfo{title}{Bulk magnetoelectricity in the
  hexagonal manganites and ferrites}}.
\newblock {\emph{\JournalTitle{Nature communications}}}
  \textbf{\bibinfo{volume}{5}}, \bibinfo{pages}{2998} (\bibinfo{year}{2014}).

\bibitem{lau2022reproducibility}
\bibinfo{author}{Lau, C.~N.}, \bibinfo{author}{Bockrath, M.~W.},
  \bibinfo{author}{Mak, K.~F.} \& \bibinfo{author}{Zhang, F.}
\newblock \bibinfo{journal}{\bibinfo{title}{Reproducibility in the fabrication
  and physics of moir{\'e} materials}}.
\newblock {\emph{\JournalTitle{Nature}}} \textbf{\bibinfo{volume}{602}},
  \bibinfo{pages}{41--50} (\bibinfo{year}{2022}).

\bibitem{basini2024terahertz}
\bibinfo{author}{Basini, M.} \emph{et~al.}
\newblock \bibinfo{journal}{\bibinfo{title}{Terahertz electric-field-driven
  dynamical multiferroicity in srtio3}}.
\newblock {\emph{\JournalTitle{Nature}}} \bibinfo{pages}{1--6}
  (\bibinfo{year}{2024}).

\bibitem{kerby2023advanced}
\bibinfo{author}{Kerby, J.}
\newblock \bibinfo{title}{The advanced photon source upgrade: A brighter future
  for x-ray science} (\bibinfo{year}{2023}).

\bibitem{fohtung2018bragg}
\bibinfo{author}{Fohtung, E.}, \bibinfo{author}{Karpov, D.} \&
  \bibinfo{author}{Baumbach, T.}
\newblock \bibinfo{journal}{\bibinfo{title}{Bragg coherent diffraction imaging
  techniques at 3rd and 4th generation light sources}}.
\newblock {\emph{\JournalTitle{Materials Discovery and Design: By Means of Data
  Science and Optimal Learning}}} \bibinfo{pages}{203--215}
  (\bibinfo{year}{2018}).

\bibitem{nazirkar2024coherent}
\bibinfo{author}{Nazirkar, N.~P.}, \bibinfo{author}{Shi, X.},
  \bibinfo{author}{Shi, J.}, \bibinfo{author}{N'Gom, M.} \&
  \bibinfo{author}{Fohtung, E.}
\newblock \bibinfo{journal}{\bibinfo{title}{Coherent diffractive imaging with
  twisted x-rays: Principles, applications, and outlook}}.
\newblock {\emph{\JournalTitle{Applied Physics Reviews}}}
  \textbf{\bibinfo{volume}{11}} (\bibinfo{year}{2024}).

\bibitem{segall2002first}
\bibinfo{author}{Segall, M.} \emph{et~al.}
\newblock \bibinfo{journal}{\bibinfo{title}{First-principles simulation: ideas,
  illustrations and the castep code}}.
\newblock {\emph{\JournalTitle{Journal of physics: condensed matter}}}
  \textbf{\bibinfo{volume}{14}}, \bibinfo{pages}{2717} (\bibinfo{year}{2002}).

\bibitem{giannozzi2009quantum}
\bibinfo{author}{Giannozzi, P.} \emph{et~al.}
\newblock \bibinfo{journal}{\bibinfo{title}{Quantum espresso: a modular and
  open-source software project for quantum simulations of materials}}.
\newblock {\emph{\JournalTitle{Journal of physics: Condensed matter}}}
  \textbf{\bibinfo{volume}{21}}, \bibinfo{pages}{395502}
  (\bibinfo{year}{2009}).

\bibitem{perdew1996generalized}
\bibinfo{author}{Perdew, J.~P.}, \bibinfo{author}{Burke, K.} \&
  \bibinfo{author}{Ernzerhof, M.}
\newblock \bibinfo{journal}{\bibinfo{title}{Generalized gradient approximation
  made simple}}.
\newblock {\emph{\JournalTitle{Physical review letters}}}
  \textbf{\bibinfo{volume}{77}}, \bibinfo{pages}{3865} (\bibinfo{year}{1996}).

\bibitem{monkhorst1976special}
\bibinfo{author}{Monkhorst, H.~J.} \& \bibinfo{author}{Pack, J.~D.}
\newblock \bibinfo{journal}{\bibinfo{title}{Special points for brillouin-zone
  integrations}}.
\newblock {\emph{\JournalTitle{Physical review B}}}
  \textbf{\bibinfo{volume}{13}}, \bibinfo{pages}{5188} (\bibinfo{year}{1976}).

\bibitem{baroni2001phonons}
\bibinfo{author}{Baroni, S.}, \bibinfo{author}{De~Gironcoli, S.},
  \bibinfo{author}{Dal~Corso, A.} \& \bibinfo{author}{Giannozzi, P.}
\newblock \bibinfo{journal}{\bibinfo{title}{Phonons and related crystal
  properties from density-functional perturbation theory}}.
\newblock {\emph{\JournalTitle{Reviews of modern Physics}}}
  \textbf{\bibinfo{volume}{73}}, \bibinfo{pages}{515} (\bibinfo{year}{2001}).

\end{thebibliography}

\noindent \textbf{For Reviews only, highlighted references (optional)} Please select 5–-10 key references and provide a single sentence for each, highlighting the significance of the work.

\section*{Acknowledgements}
E.F. and M.N conceived the project and conceptualized the experiments. J.S. and J.J. grew and provided free-standing and epitaxial 2D CBNO samples. N.N, V.T, J.B, W.C,R.H and E.F Performed the OAM BCDI measurements and characterization. BCDI analysis using phase retrieval algorithms was performed by N.N and E.F OAM Raman measurements and analysis was performed by N.N, V.T, P.B, A.N, M.N, and E.F., A.D performed the DFT computations. N.N and E.F. wrote the paper. All the authors discussed the results and commented on the manuscript. 
We thank S. Williams (RPI),  A. Ndao (UCSD), L. Caretta (Brown University), and D. Kumah (Duke University) for the stimulating conversations on this manuscript.

\textbf{Funding:}
This work was supported by the US Department of Energy (DOE), Basic Energy Sciences, Materials Sciences and Engineering Division, under grant No. DE-SC0023148 (OBCDI technique, ORAMAN and analysis), and the US Department of Defense, Air Force Office of Scientific Research (AFOSR), under award No. FA9550-23-1-0325 (Program Manager: Dr. Ali Sayir) for work on probing topological vortices and piezoelectric enhancements. J.S. thanks the support from NSF under award number 2024972. This research used resources of the Advanced Photon Source (APS), a U.S. Department of Energy (DOE) Office of Science User Facility, operated for the DOE Office of Science by Argonne National Laboratory (ANL) under contract No. DE-AC02-06CH11357.~\textbf{Author contributions:} R.H., W.C., and E.F. designed and performed the BCDI experiment. J.J. and J.S. grew and provided the samples. V.T, P.B., and M.N synthesised the twisted beam.M.N and V.T acknowledge the support of The Office of Advanced Scientific Computing Research within the  Office of Science US Department of Energy (DOE) under grant No DE-SC0024676. All authors interpreted the results and contributed to
writing the manuscript.~\textbf{Competing interests:} The authors declare no competing interests.~\textbf{Data and materials availability:} The data reported in this paper are archived at Beamline 34-ID-C at the Advanced Photon Source at Argonne National Lab and are
available upon request. All code, including the reconstruction
algorithm, is also available upon request.

    
\section*{Data availability}
The experimental and DFT data are available upon reasonable request.

\section*{Competing interests}
The authors declare no competing interests.

\section*{Publisher’s note}
Springer Nature remains neutral with regard to jurisdictional claims in published maps and institutional affiliations.

\section*{Supplementary information (optional)}

\end{document}